\documentclass[fleqn,twoside]{article}
\usepackage{espcrc2}
\usepackage{epsfig}

\def\la{\mathrel{\mathchoice {\vcenter{\offinterlineskip\halign{\hfil
$\displaystyle##$\hfil\cr<\cr\sim\cr}}}
{\vcenter{\offinterlineskip\halign{\hfil$\textstyle##$\hfil\cr<\cr\sim\cr}}}
{\vcenter{\offinterlineskip\halign{\hfil$\scriptstyle##$\hfil\cr<\cr\sim\cr}}}
{\vcenter{\offinterlineskip\halign{\hfil$\scriptscriptstyle##$\hfil\cr<\cr
\sim\cr}}}}}

\def\ga{\mathrel{\mathchoice {\vcenter{\offinterlineskip\halign{\hfil
$\displaystyle##$\hfil\cr>\cr\sim\cr}}}
{\vcenter{\offinterlineskip\halign{\hfil$\textstyle##$\hfil\cr>\cr\sim\cr}}}
{\vcenter{\offinterlineskip\halign{\hfil$\scriptstyle##$\hfil\cr>\cr\sim\cr}}}
{\vcenter{\offinterlineskip\halign{\hfil$\scriptscriptstyle##$\hfil\cr>\cr
\sim\cr}}}}}
\newcommand{\etal}{{\it et al.\/}}

\newcommand{\AmS}{{\protect\the\textfont2
  A\kern-.1667em\lower.5ex\hbox{M}\kern-.125emS}}

\title{Cosmic Ray Energy Spectra and Mass Composition at the 
Knee\\ -- Recent Results from KASCADE --}

\author{
K.-H.~Kampert \address[BUW]{ Fachbereich C - Physik,
Bergische Universit\"at Wuppertal, 42097 Wuppertal, Germany}
\thanks{\tt email: kampert@uni-wuppertal.de},
T.~Antoni \address[IEKP]{ Institut f\"ur Experimentelle Kernphysik,
Universit\"at Karlsruhe, 76021 Karlsruhe, Germany},
W.D.~Apel \address[IK]{ Institut\ f\"ur Kernphysik, Forschungszentrum Karlsruhe,
76021~Karlsruhe, Germany}, 
F.~Badea \addressmark[IK],
K.~Bekk \addressmark[IK], 
A.~Bercuci \address[IPNE]{ National Institute of Physics and Nuclear
Engineering, 7690~Bucharest, Romania},
H.~Bl\"umer \addressmark[IEKP]\addressmark[IK],
H.~Bozdog \addressmark[IK],
I.M.~Brancus \addressmark[IPNE],
C.~B\"uttner \addressmark[IEKP],
A.~Chilingarian \address[Yere]{ Cosmic Ray Division, Yerevan Physics
Institute, Yerevan~36, Armenia},
K.~Daumiller \addressmark[IEKP],
P.~Doll \addressmark[IK],
R.~Engel \addressmark[IK],
J.~Engler \addressmark[IK],
F.~Fe{\ss}ler \addressmark[IEKP],
H.J.~Gils \addressmark[IK],
R.~Glasstetter \addressmark[BUW],
A.~Haungs \addressmark[IK],
D.~Heck \addressmark[IK],
J.R.~H\"orandel \addressmark[IEKP],
H.O.~Klages \addressmark[IK],
G.~Maier \addressmark[IK],
H.J.~Mathes \addressmark[IK],
H.J.~Mayer \addressmark[IK],
J.~Milke \addressmark[IK],
M.~M\"uller \addressmark[IK],
R.~Obenland \addressmark[IK],
J.~Oehlschl\"ager \addressmark[IK],
S.~Ostapchenko \addressmark[IEKP],
M.~Petcu \addressmark[IPNE],
S.~Plewnia \addressmark[IK],
H.~Rebel \addressmark[IK],
A.~Risse \address[Soltan]{ Soltan Institute for Nuclear Studies,
90950~Lodz, Poland},
M.~Risse \addressmark[IK],
M.~Roth \addressmark[IEKP],
G.~Schatz \addressmark[IK],
H.~Schieler \addressmark[IK],
J.~Scholz \addressmark[IK],
M.~St\"umpert \addressmark[IEKP],
T.~Thouw \addressmark[IK],
H.~Ulrich \addressmark[IK],
J.~van~Buren~\addressmark[IK],
A.~Vardanyan \addressmark[Yere],
A.~Weindl \addressmark[IK],
J.~Wochele \addressmark[IK],
J.~Zabierowski \addressmark[Soltan],
S.~Zagromski \addressmark[IK]
}

\begin{document}

\begin{abstract}
Recent results from the KASCADE experiment on measurements of
cosmic rays in the energy range of the knee are presented.
Emphasis is placed on energy spectra of individual mass groups as
obtained from an two-dimensional unfolding applied to the
reconstructed electron and truncated muon numbers of each
individual EAS. The data show a knee-like structure in the energy
spectra of light primaries (p, He, C) and an increasing dominance
of heavy ones ($A \ga 20$) towards higher energies.  This basic
result is robust against uncertainties of the applied interaction
models QGSJET and SIBYLL which are used in the shower 
simulations to analyse the data. Slight differences observed between
experimental data and EAS simulations provide important clues for
further improvements of the interaction models.  The data are
complemented by new limits on global anisotropies in the arrival
directions of CRs and by upper limits on point sources.
Astrophysical implications for discriminating models of maximum
acceleration energy vs galactic diffusion/drift models of the
knee are discussed based on this data.  \vspace{1pc}
\end{abstract}

\maketitle

\section{Introduction}
\begin{figure}[t]
\centerline{\epsfxsize=\columnwidth\epsfbox{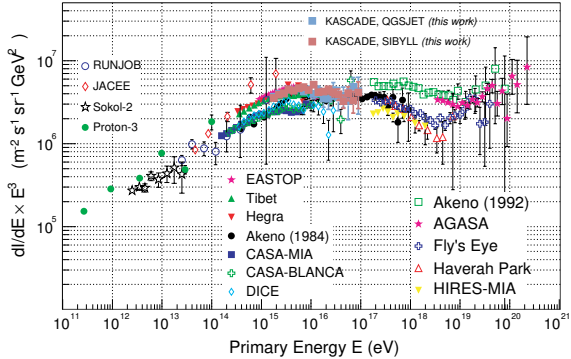}}
\vspace*{-5mm} \caption[]{Compilation of the all-particle cosmic
ray spectrum showing the knee, the suggested second knee, and the
ankle of the CR spectrum (compilation by H. Ulrich).
\label{fig:all-particle-compilation}}
\end{figure}

A puzzling and most prominent feature of the cosmic ray (CR)
spectrum is the so-called knee, where the spectral index of the
all-particle power-law spectrum changes from approximately $-2.7$
to $-3.1$.  Several models have been proposed in order to explain
this feature shown in Fig.\,\ref{fig:all-particle-compilation},
but none of them has managed to become broadly accepted.  Some
models focus on a possible change in the acceleration mechanism
at the knee~\cite{Lagage83,Biermann93,Drury94}, e.g.\ due to the
limiting energy defined by the size and magnetic field strength
of the acceleration region ($E_{\rm max} \la Z \times (B \times
L))$.  Other ones discuss an increased leakage of CRs from the
Galaxy due to a change in the confinement efficiency by galactic
magnetic fields, e.g.~\cite{Wdowczyk84,Ptuskin93,Candia02a,Roulet03}.
Again, this results in a rigidity scaling of the knee according
to the maximum confinement energy.  Finally, a third group of
models attributes the effect of the knee to CR interactions at
their sources, during their propagation in the Galaxy, or in the
upper atmosphere.  Such scenarios include nuclear
photo-disintegration processes by UV-photons at the
sources~\cite{Candia02b}, interactions of CRs in dense fields of
massive relic neutrinos~\cite{Wigmans03}, production of gravitons
in high-energy pp collisions~\cite{Kazanas01}, etc.  A recent
review about this topic can be found e.g.\ in
Refs.~\cite{Kampert01a,Haungs03}.

To distinguish between these models and allowing to answer the
long pressing question about the origin of cosmic rays and about
the knee in their spectrum, high quality and high statistics data
are required over an energy interval ranging from at least 0.5 to
500 PeV. It appears worthwhile to mention that solving the old
problem about the origin of CRs in the PeV region is a
prerequisite also for an understanding of the highest energies in
the GZK-region.  Due to the low flux involved at energies $\ga
10^{14}$ eV, only extensive air shower (EAS) experiments are able
to provide such data.  In EAS experiments, primary CRs are only
indirectly observed via their secondaries generated in the
atmosphere.  The most important experimental observables at
ground are then the electromagnetic (electrons and photons),
muonic, and hadronic components.  In addition or alternatively,
some experiments also detect photons originating from Cherenkov
and/or fluorescence radiation of charged particles in the
atmosphere.  For a brief review about EAS observables and their
experimental techniques the reader is referred to
Refs.\,\cite{Kampert01b,Swordy02,Haungs03}.

Unfortunately, progress on interpreting EAS data has been 
modest mostly because of two reasons: Firstly, the EAS development
is driven both by the poorly known high-energy hadronic
interactions and their particle production in the very forward
kinematical region as well as by uncertainties in the low energy
interaction models influencing mostly the lateral particle
density distribution functions \cite{Drescher04}.  Secondly, due
to the stochastic nature of particle interactions, most
importantly the height of the very first interaction in the
atmosphere, EAS are subject to large fluctuations in particle
numbers at ground.  To make things even more complicated, the
amount of fluctuations depends, amongst others, sensitively on
the primary CR energy and mass~\cite{Kampert01b}.  Here, it is
very important to realize that EAS fluctuations are not to be
mistaken as random Gaussian errors associated with the statistics
in the number of particles observed at ground.  The latter one
can be improved by the sampling area of an EAS experiment, while
the former one is intrinsic to the EAS itself, carrying - for a
sample of events - important information about the nature of the
primary particle.  Clearly, both kinds of fluctuations have to be
accounted for in the data analysis of steeply falling energy
spectra in order to avoid misinterpretations of the observations.

\section{Results from the KASCADE Experiment}
KASCADE (\underline{Ka}rlsruhe \underline{S}hower
\underline{C}ore and \underline{A}rray \underline{De}tector) is a
sophisticated EAS experiment for detailed investigations of
primary CRs in the energy range of the knee.  For reconstructing
the CR energy and mass and for investigating high-energy hadronic
interactions, KASCADE follows the concept of a multi-detector
set-up providing as much complementary information as possible as
well as redundancy for consistency tests.  Most relevant for the
results presented in this paper is the scintillator array
comprising 252 detector stations of electron and muon counters
arranged on a grid of $200 \times 200$ m$^{2}$.  In total, it
provides about 500 m$^2$ of $e/\gamma$- and 620 m$^{2}$ of
$\mu$-detector coverage.  The detection thresholds for vertical
incidence are $E_{e}^{\rm thr} \simeq 5$ MeV and $E_{\mu}^{\rm
thr} \simeq 230$ MeV. More details about the $e/\gamma$- and
$\mu$-detectors and all other detector components can be found in
Ref.\,\cite{KASCADE-NIM}.

\subsection{Chemical Composition and Energy Spectra}

\begin{figure}[t]
\centerline{\epsfxsize=\columnwidth\epsfbox{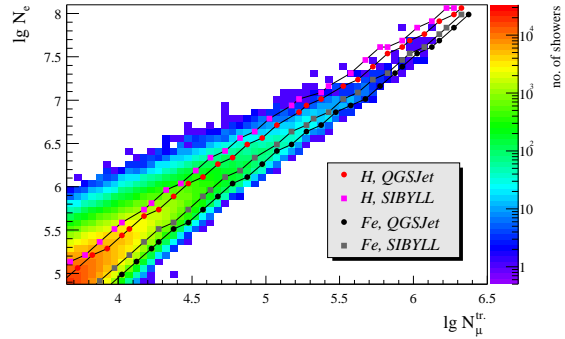}}
\vspace*{-5mm} \caption[]{Two dimensional electron ($N_{e}$) and
truncated muon number ($N_{\mu}^{\rm tr}=\int_{40\rm{ m}}^{200\rm{ m}}
\rho_{\mu}(r) dr$) spectrum measured by
the KASCADE array.  Lines display the most probable values
expected for proton and iron primaries according to CORSIKA
simulations employing two different hadronic interaction
models~\cite{Ulrich03}.
\label{fig:2d}}
\end{figure}

The traditional and perhaps most sensitive technique to infer the
CR composition from EAS data is based on measurements of the
electron ($N_e$) and muon numbers ($N_\mu$) at ground.  It is
well known~\cite{Kampert01b} that for given energy, primary
Fe-nuclei result in more muons and fewer electrons at ground as
compared to proton primaries.  Specifically, in the energy range
and at the atmospheric depth of KASCADE, a Fe-primary yields
about 30\,\% more muons and almost a factor of two fewer
electrons as compared to a proton primary.  The basic
quantitative procedure of KASCADE for obtaining the energy and
mass of the cosmic rays is a technique of unfolding the observed
two-dimensional electron vs truncated muon number spectrum of
Fig.\,\ref{fig:2d} into the energy spectra of primary mass
groups~\cite{Ulrich03}.  The problem can be considered a system
of coupled Fredholm integral equations of the form
\begin{eqnarray*}
\lefteqn{\frac{dJ}{d\,\lg N_e \;\; d\,\lg N_\mu^{\rm tr}} =
\sum_A \int\limits_{-\infty}^{+\infty} \frac{d\,J_A}{d\,\lg E} 
\quad \cdot} \\
& &
\cdot \quad p_A(\lg N_e\, , \,\lg N_\mu^{\rm tr}\, \mid \, \lg E)
  \cdot d\, \lg E
\end{eqnarray*}
where the probability $p_A$
\begin{eqnarray*}
\lefteqn{p_A(\lg N_e , \lg N_\mu^{\rm tr}\, \mid \, \lg 
    E) =} \\
& & \int\limits_{-\infty}^{+\infty} k_A(\lg N_e^t , \lg 
    N_\mu^{\rm tr,t})
  d\, \lg N_e^t\,\, d\,\lg N_\mu^{\rm tr,t}
\end{eqnarray*}
is another integral equation with the kernel function $k_A = r_A
\cdot \epsilon_A \cdot s_A$ factorizing into three parts.  Here,
$r_A$ describes the shower fluctuations, i.e.\ the 2-dim
distribution of electron and truncated muon number for fixed
primary energy and mass, $\epsilon_A$ describes the trigger
efficiency of the experiment, and $s_A$ the reconstruction
probabilities, i.e.\ the distribution of $N_e$ and $N_\mu^{\rm
tr}$ that is reconstructed for given {\em true} numbers $N_e^t$,
$N_\mu^{\rm tr,t}$ of electron and truncated muon numbers.  The
probabilities $p_A$ are obtained by parameterizations of EAS
Monte Carlo simulations for fixed energies using a moderate
thinning procedure as well as smaller samples of fully simulated
showers for the input of the detector simulations.  Because of
the shower fluctuations mentioned above, unfolding of all 26
energy spectra ranging from protons to Fe-nuclei is clearly
impossible.  Therefore, 5 elements (p, He, C, Si, Fe) were chosen
as representatives for the entire distribution.  More mass groups
do not improve the $\chi^2$-uncertainties of the unfolding but
may result in mutual systematic biases of the reconstructed
spectra \cite{Ulrich03}.

The unfolding procedure is tested by using random initial spectra
generated by Monte Carlo simulations.  It has been
shown~\cite{Ulrich03} that knee positions and slopes of the
initial spectra are well reproduced and that the discrimination
between the five primary mass groups is sufficient.  For
scrutinizing the unfolding procedure, different mathematical ways
of unfolding (Gold-algorithm, Bayes analyses, principle of
maximum entropy, etc.)  have been compared and the results are
consistent~\cite{Ulrich03}.  For generating the kernel functions
a large number of EAS has been
simulated~\cite{Ulrich03,roth-ulrich} employing
CORSIKA~\cite{cors} with the hadronic interaction models QGSJET
(version 2001)~\cite{qgs} and SIBYLL 2.1~\cite{sib}.

\begin{figure}[t]
\centerline{\epsfxsize=\columnwidth\epsfbox{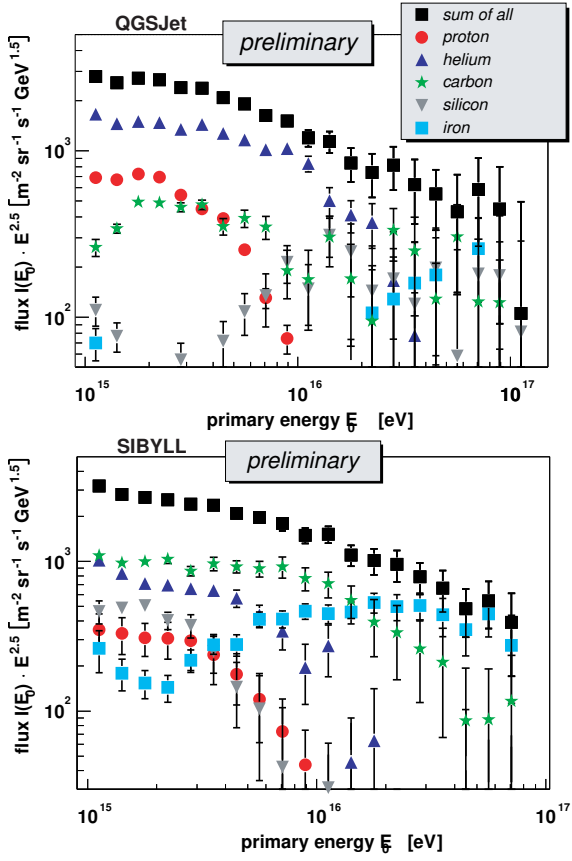}}
\vspace*{-5mm}
\caption[]{Results of the unfolding procedure using QGSJET (left) 
and SIBYLL (right) as hadronic interaction model~\cite{Ulrich03}.
\label{fig:spectra}}
\end{figure}

The result of the unfolding is presented in
Fig.\,\ref{fig:spectra} for each of the two interaction models.
Clearly, there are common features but also differences in the
energy distributions obtained with the two interaction models.
The all-particle spectra, also shown in 
Fig.~\ref{fig:all-particle-compilation},
coincide very nicely and in both cases
the knee is caused by the decreasing flux of the light primaries,
corroborating results of an independent analysis of
Ref.\,\cite{muon-density}.  Tests using different data sets,
different unfolding methods, etc.\ show the same
behavior~\cite{roth-ulrich}.  As the most striking difference,
SIBYLL suggests a more prominent contribution of heavy primaries
at high energies.  This difference results from the different
$N_{e}$-$N_{\mu}^{\rm tr}$ correlation shown in
Fig.\,\ref{fig:2d}, i.e.\ SIBYLL predicts higher electron and
lower muon numbers for given primaries as compared to QGSJET.

\begin{figure}[t]
\centerline{\epsfxsize=\columnwidth\epsfbox{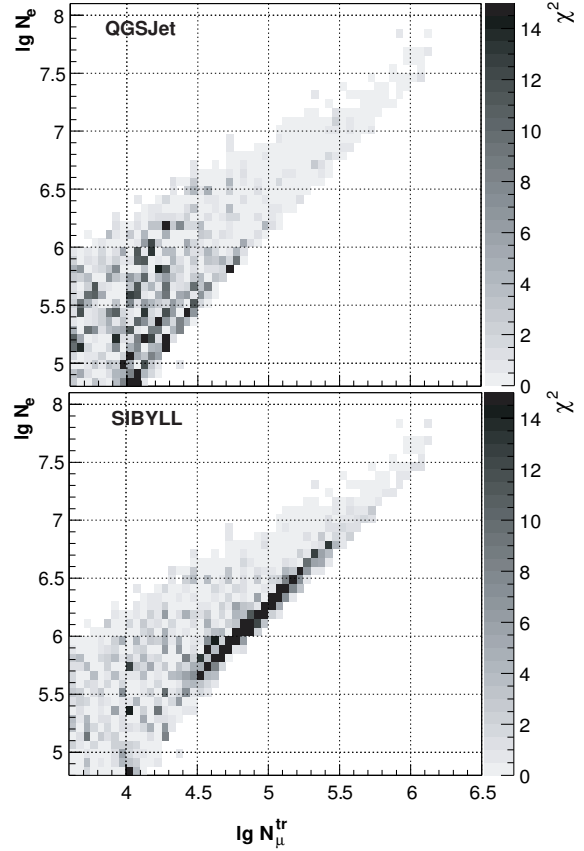}}
\vspace*{-5mm}
\caption[]{$\chi^{2}$ deviation between the forward folded 
reconstructed and measured  
($N_{e}$-$N_{\mu}^{\rm tr}$)-data cells for the QGSJET (left) 
and SIBYLL (right) hadronic interaction models~\cite{Ulrich03}.
\label{fig:chi2}}
\end{figure}

Is there a way to judge which of the two models describes the
data better?  This is done most easily by comparing the residuals
of the unfolded two-dimensional $N_{e}$ vs $N_{\mu}^{\rm tr}$
distributions with the actual data used as input to the unfolding
(Fig.\,\ref{fig:2d}).  The result of such an analysis is
presented in Fig.\,\ref{fig:chi2} in terms of $\chi^2$.  The
deviations reveal some deficiencies of QGSJET at low electron and
muon numbers and they clearly demonstrate that SIBYLL encounters
problems in describing the high-$N_{e}$ - low-$N_{\mu}^{\rm tr}$
tail of the experimental data at 10 PeV and
above~\cite{Ulrich03}.  If not being prepared to accept an
additional significant contribution of superheavy primaries ($A >
60$) required in case of SIBYLL simulations to fill the gap at
high muon numbers, the results suggest a muon deficit (a/o
electron excess) in this model.  Definitely, this problem
needs further attention and will be very important also for
composition studies at higher energies~\cite{Watson04}.

With this caveats kept in mind, the KASCADE data favor an
astrophysical interpretation of the knee and are in agreement
with a constant rigidity of the knee position for the different
primaries.  Similar results were very recently obtained for an
analysis of two-mass groups based on combined EAS-TOP / MACRO
measurements~\cite{Aglietta03}, and were again confirmed for
three mass groups from EAS-TOP electron and muon
measurements~\cite{Navarra04}.  Within the given error bars, the
mean logarithmic masses of both experiments agree well with one
another.

\subsection{Search for Anisotropies and Point Sources}

Additional information about the origin of CRs and their
propagation in the galactic environment can be obtained from
global anisotropies in their arrival directions.  Model
calculations show that diffusion of CRs in the galactic magnetic
field can result in anisotropies on a scale of $10^{-4}$ to
$10^{-2}$ depending on the energy of the particle, the strength
and structure of the galactic magnetic field \cite{Candia02a},
and on the source distribution.  Since the diffusion scales again
with the rigidity, a factor of 5-10 larger anisotropies are
expected for protons as compared to iron primaries.  This
rigidity dependent diffusion is one of the possible explanations
of the knee.

\begin{figure}[t]
\centerline{\epsfxsize=\columnwidth\epsfbox{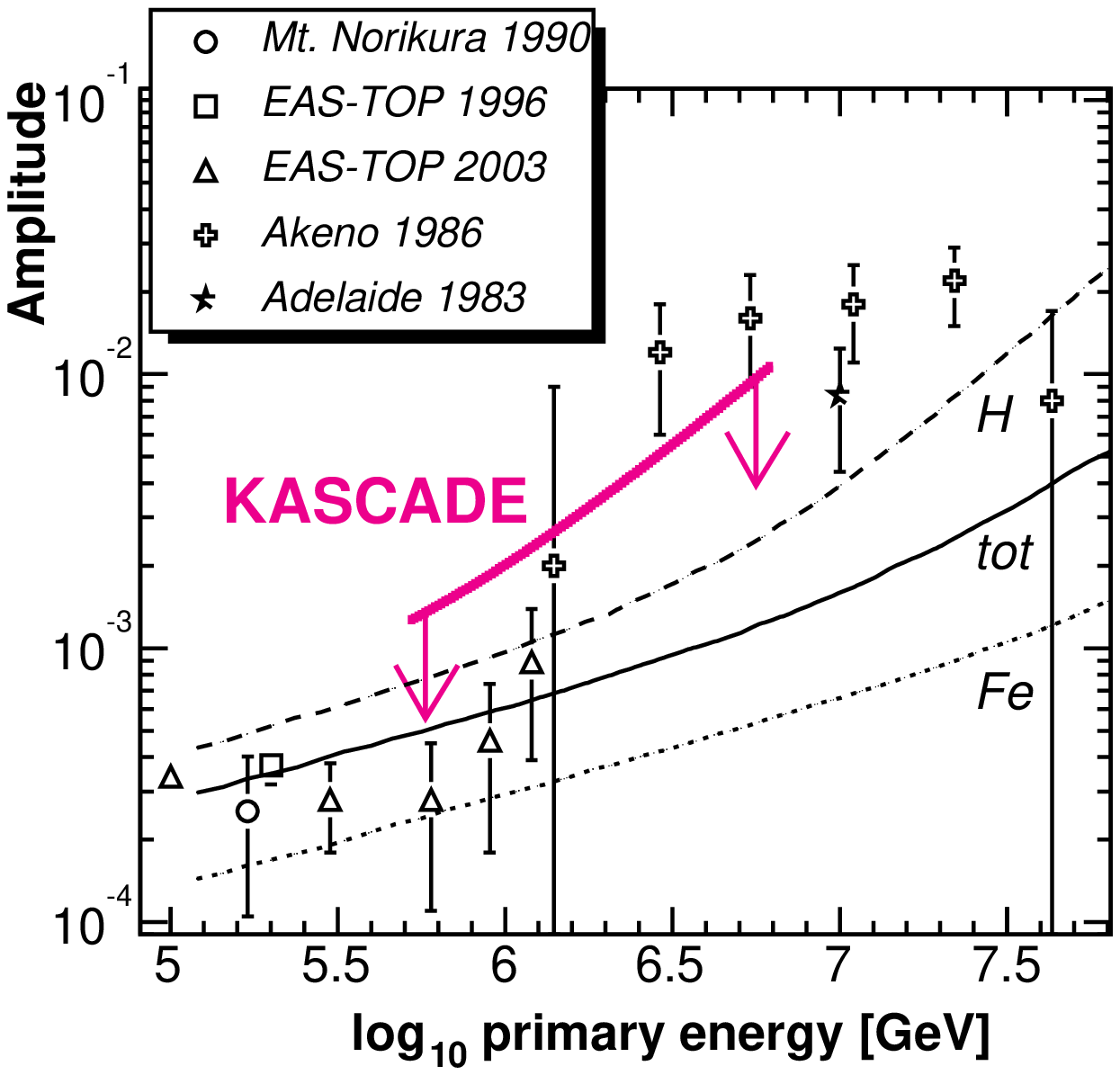}}
\vspace*{-5mm}
\caption[]{KASCADE upper limits (95\,\%) of Rayleigh
amplitudes $A$ vs primary energy (bold line) compared to results
from other experiments and to expectations from galactic CR
diffusion (thin lines)~\cite{global-anisotropies}.
\label{fig:global}}
\end{figure}

\begin{figure}[t]
\centerline{\epsfxsize=\columnwidth\epsfbox{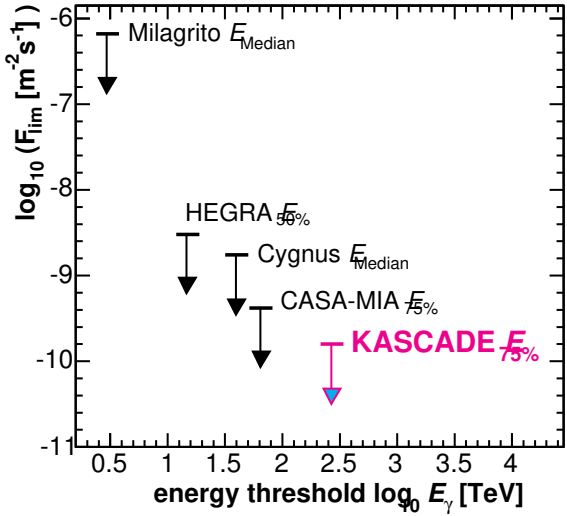}}
\vspace*{-5mm}
\caption[]{90\,\% upper limit for a point source moving through the
zenith in comparison with results from other
experiments~\cite{point-sources}.  Note the different definitions
of the energy thresholds.
\label{fig:point}}
\end{figure}

Because of the small anisotropy expected, a large data sample and
careful data selection is necessary.  About $10^{8}$ EAS events
in the energy range from 0.7 to 6 PeV were selected and studied
in terms of Rayleigh amplitudes $A$ and phases $\Phi$ of the
first harmonic.  Neither for the full set of data nor for
electron-rich and -poor EAS significant Rayleigh amplitudes were
found.  The upper limit on the large scale anisotropy is depicted
in Fig.\,\ref{fig:global} \cite{global-anisotropies} and is in
line with results reported from other experiments.  We shall come
back to this result in the next section.

Even though the location of CR sources should be obscured due to
the deflection of charged particles in the magnetic field of our
galaxy, there is interest to perform point source searches.  For
example, neutrons are not deflected and can reach the Earth if
their energy and hence decay length is comparable with the
distance of the source.  A decay length of 1 kpc corresponds to a
neutron energy of about $10^{17}$ eV. Also, by applying
appropriate cuts to electron and muon numbers from EAS, searches
for $\gamma$-ray point sources can be performed in the PeV range.

Such a study has been performed based on 47 Mio EAS with primary
energies above $\sim 300$ TeV. A certain region in the sky is
then analyzed by comparing the number of events from the assumed
direction with an expected number of background events.  For the
latter, the so-called time-shuffling method has been used.  As a
result, again no significant excess is found in the region of the
galactic plane or for selected point source candidates.  Assuming
equal power laws in the energy spectra of background and source
events, upper flux limits can be calculated for given energy
thresholds.  For a steady point source that transits the zenith,
we obtain an upper flux limit of $3 \cdot 10^{-10}$
m$^{-2}$s$^{-1}$ (see Fig.\,\ref{fig:point})
\cite{point-sources}. This is roughly 1-2 orders of magnitude 
larger than the Crab flux extrapolated to this energy.

Very recently, Chilingarian \etal ~reported the detection of a
source of high-energy CRs in the Monogem ring
\cite{chili-source}.  Changing slightly our cuts in zenith angle
to widen the declination range thereby covering the position of
the source candidate, we find 742 events within an opening angle
of $0.5^{\circ}$ around the suggested location with an expected
number of 716 background events yielding an upper flux limit of
$3 \cdot 10^{-10}$ m$^{-2}$s$^{-1}$.  Similar values are found
when searching for an excess from the direction of the pulsar PSR
B0656+14 \cite{Thorsett03} located near the centre of the Monogem
SNR.

\subsection{Implications for understanding the CR origin}

The new high quality data presented in the previous sections have
revitalized the interest to understand both the origin of CRs in
the knee region and the phenomenon of the knee structure itself.

This is because discriminating models of maximum acceleration
energy from galactic diffusion/drift models of the knee or from
particle physics interpretations require detailed inspection of
knee structures seen in {\em individual} mass groups combined
with precise measurements of CR anisotropies.  Of particular
interest are the energies of the spectral breaks, the power-law
indices below and above the knee, and the smoothness of the
turn-over regions.  Even though, these goals are not yet achieved
totally, important steps have been made towards it.  Previous
investigations were limited to inclusive CR all-particle spectra
and to global changes of the mean logarithmic mass, $\ln A$, with
primary energy.  Such measurements appeared to be too insensitive
for a convincing discrimination of models.  Furthermore, in most
cases not full distributions but only {\em mean} values of
experimental and simulated distributions were compared to each
other.  A prominent example are plots of the shower maximum
$X_{max}$ vs primary energy.  Obviously, deficiencies of hadronic
interaction models remain unrecognized in such plots, unless EAS
data are below proton or above iron simulations.

\begin{figure}[t]
\centerline{\epsfxsize=\columnwidth\epsfbox{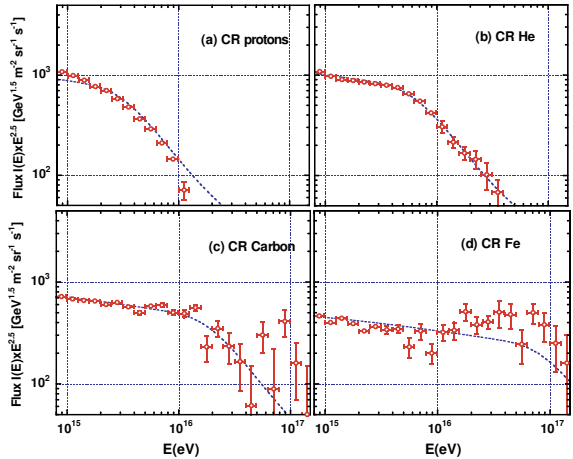}}
\vspace*{-5mm}
\caption[]{Fit of the GRB-model of Wick \etal ~\cite{Wick04} to
the preliminary KASCADE data presented at ICRC 2001
\cite{Kampert01c,Ulrich01}.  In the model, a GRB occurred
210\,000 years ago at a distance of 500 pc and injected
$10^{52}$~ergs into CRs. The CRs isotropically diffuse with an
energy-dependent mean-free-path in a MHD turbulence field.
\label{fig:wick-spectra}}
\end{figure}

Very good examples demonstrating the discrimination power of the
new data presented here and showing the amount of information
contained in it is given by recent studies of Wick
\etal~\cite{Wick04} or Dar~\cite{Dar04}.  Based on the earlier
suggested connection between Gamma-Ray Bursts (GRBs) and
ultrahigh-energy CRs \cite{Vietri95,Waxman95} they propose a
model for the origin of CRs from $\sim 10^{14}$~eV/nucleon up to
the highest energies ($\ga 10^{20}$~eV).  In that model, GRBs are
assumed to inject CR protons and ions into the interstellar
medium of star-forming galaxies - including the Milky Way - with
a power-law spectrum extending to a maximum energy $\sim
10^{20}$~eV. High-energy CRs injected in the Milky Way diffuse
and escape from our Galaxy.  Ultra high-energy CRs with energies
$\ga 10^{17}$ to $10^{18}$\,eV that have Larmor radii comparable
to the size scale of the galactic halo escape directly from the
Milky Way and propagate almost rectilinearly through
extragalactic space.  By the same token, UHECRs produced from
other galaxies can enter the Milky Way to be detected.  UHECRs
formed in GRBs throughout the universe then travel over
cosmological distances and have their spectrum modified by energy
losses, so an observer in the Milky Way will measure a
superposition of UHECRs from extragalactic GRBs and HECRs
produced in our Galaxy.

Thereby, the CR spectrum near the knee is understood by CRs
trapped in the Galactic halo that were accelerated and injected
by an earlier Galactic GRB. Assuming magneto-hydrodynamical
turbulence superposed to the galactic magnetic field, a fit of
the model of Ref.\,~\cite{Wick04} to the preliminary KASCADE
data, shown in Fig.\ \ref{fig:wick-spectra}, suggests a 500 pc
distant GRB that released $10^{52}$~ergs in CRs if the GRB took
place about 210\,000 yrs ago.  Keeping in mind the still large
uncertainties of the data and some freedom of parameters in the
model, there is remarkable accordance observed.  In this model,
the rigidity dependence of knee position is caused by galactic
modulation effects.  The GRB-model of Dar~\cite{Dar04}, on the
other hand, predicts the knee to be proportional to the {\em
mass} $A$ of CR primaries.  This is because of the relativistic
beaming effect (cannon balls) in the SN-jets of that model.  It
also fits the data within the present uncertainties with
differences showing up mostly for protons and Helium ($Z/A=1,
0.5$, respectively).

Since in these models very few or just a single galactic GRB is
responsible for most of the CRs in the knee region, anisotropies
in the arrival directions are expected on different levels,
depending on the distance and age of the GRB. For example, the
authors of Ref.~\cite{Wick04} state that if an anisotropy below
$\sim 0.2$\,\% is confirmed, then a number of implications
follow.  Either we are located near a rather recent GRB, which
could be unlikely, or the CR energy release from GRBs is larger
than the one given above \cite{Wick04}.  Thus, improving the
anisotropy limits of the previous section would help to further
pin down this model.  Similarly, recent calculations of standard
SN acceleration models show that global anisotropies well in
excess even of 0.2\,\% are expected for more realistic source
distributions in the Galaxy, dependent on the structure of the
magnetic fields \cite{Ptuskin-ICRC}.  The upper limits provided by
the present data rule out already a large parameter space in
these calculations.

However, before starting to over-interpret the data, particularly
the energy spectra of mass groups, we should emphasize again
their sensitivity to the applied interaction models.  More work
is still needed to improve the models and to arrive at smaller
systematic uncertainties.  On the other hand, the very valuable
study of such models of CR origin demonstrates the informational
content reached by present data.

\section{Summary and Outlook}

KASCADE has provided a wealth of new high quality EAS data in the
knee region giving important insight into the origin of the knee
and of CRs in general.  Conclusive evidence has been reached on
the knee being caused by light primaries (p + He) mostly.
Furthermore, the data are in agreement with a rigidity scaling of
the knee position giving support to an astrophysical origin by
either maximum acceleration or diffusion/drift models of
propagation.  Astrophysical parameters start to be constrained by
the preliminary KASCADE data, as was demonstrated at the example
of the GRB models of Refs.\,\cite{Wick04,Dar04}.  Further
important constrains result from measurements of large scale
anisotropies of CRs and by new limits on point sources.  Claims
that were made in Ref.~\cite{chili-source} about an excess of CRs
from the Monogem SNR cannot be confirmed.

A particle physics interpretation of the knee appears to be
excluded with a high level of confidence.  For example,
interactions of CRs with background particles or
photo-disintegration in the Galaxy would produce an abundance of
secondary protons and result in a light mass composition above
the knee energy, a result which is in contradiction to the
present data.  Furthermore, $\bar{\nu_{e}}$'s with a mass of
$\simeq 0.5$~eV/$c^{2}$ as are needed in the model of
Ref.~\cite{Wigmans03} appear to be excluded by recent WMAP and
2dF data for neutrinos in case of degenerated masses, with the
latter being suggested by recent oscillation and possibly by a
neutrinoless double-beta experiment (for a recent discussion of
this topic the reader is referred to
Refs.~\cite{Hannestad04},\cite{Fogli04}).

Presently, more data and more observables are being analyzed
within KASCADE, particularly in terms of composition analyses
employing reconstructions of the muon production height
\cite{Buettner03}.  Together with measurements of energetic
hadrons in the central calorimeter, the unfolding technique of
electron and muon numbers in EAS has become a powerful tool to
reconstruct the properties of primary particles in EAS and it
also provides important clues on how to improve the hadronic
interaction models employed in CORSIKA air shower simulations.

KASCADE-Grande has just started its routinely data taking and
will extend the measurements up to $10^{18}$\,eV, thereby
allowing to verify the existence of the putative Iron knee
marking the so-called second knee in the all-particle CR spectrum
\cite{Kampert03,Badea04}.  This, together with improved
statistics for anisotropy measurements will allow to confront
astro- and particle-physics motivated models of the knee in much
more detail to the experimental data as has been possible up to
now.

Furthermore, the use of radio antennas complementing the
experimental KASCADE-Grande set-up may open a new window to
future EAS observations on large scales
\cite{Horneffer03,Badea04}.  Interesting first observations have
been made and are presently being analyzed in detail.

\vspace*{5mm} {\small {\bf Acknowledgement:} It is a pleasure to
thank the organizers of the meeting for financial support and for
their invitation to participate in a very interesting and
fruitful meeting conducted in a pleasant atmosphere.  This
work is supported by Forschungszentrum Karlsruhe, the German
Ministry for Research and Education (Grant 05 CU1VK1/9), and by a
Polish KBN grant for the years 2004-6.}

\end{document}